\documentclass{article}
\usepackage{dcase2020,amsmath,graphicx,url,times,booktabs,tabularx}
\usepackage{indentfirst}
\usepackage{siunitx}
\usepackage{bm}
\usepackage[all]{nowidow}

\newcommand{\norm}[1]{\left\lVert#1\right\rVert}

\title{ID-Conditioned Auto-Encoder for Unsupervised Anomaly Detection}

\name{S\l{}awomir Kapka}
\address{Samsung R\&D Institute Poland\\
      Artificial Intelligence\\
      Warsaw, Poland \\
      s.kapka@samsung.com}

\begin{document}

\ninept
\maketitle

\begin{sloppy}

\begin{abstract}
\indent \indent
In this paper, we introduce ID-Conditioned Auto-Encoder for unsupervised anomaly detection. Our method is an adaptation of the Class-Conditioned Auto-Encoder (C2AE) designed for the open-set recognition. Assuming that non-anomalous samples constitute of distinct IDs, we apply Conditioned Auto-Encoder with labels provided by these IDs. Opposed to C2AE, our approach omits the classification subtask and reduces the learning process to the single run. We simplify the learning process further by fixing a constant vector as the target for non-matching labels. We apply our method in the context of sounds for machine condition monitoring. We evaluate our method on the ToyADMOS and MIMII datasets from the DCASE 2020 Challenge Task 2. We conduct an ablation study to indicate which steps of our method influences results the most.
\end{abstract}


\section{Introduction}
\label{sec:intro}

Unsupervised anomaly detection is a problem of detecting anomalous samples under the condition that only non-anomalous (normal) samples have been provided during training phase. In this paper, we focus on unsupervised anomaly detection in the context of sounds for machine condition monitoring -- i.e., detecting mechanical failure by listening.

Many techniques have been studied for detecting anomalous sounds. Among others there are solutions based on SVMs \cite{6027306, 7057083}, sparse coding \cite{8521320, Giri2019}, GMMs \cite{5283337, ProbabilisticNoveltyDetection}, Neyman-Pearson lemma \cite{8081297, 8501554}, signal processing \cite{8683702, 6327988, 6309608}, Interpolation DNN \cite{9054344}, and Auto-Encoders \cite{8501554, 8168164, Oh_2018, 8937183}. Beyond sounds, much more techniques for anomaly detection based on deep leaning can be found in the survey \cite{chalapathy2019deep}.

Unsupervised anomaly detection can be viewed as a special case of the open-set recognition \cite{OpenSetRecognition}. In fact, since during training we are provided only with normal samples and we have to predict whether new samples are normal or anomalous, we can look at it as a binary classification problem with only one given class during training phase. 

Recently, Class-Conditioned Auto-Encoder (C2AE) \cite{C2AE} has been introduced for the open-set recognition problem. According to the survey on the open-set recognition \cite{OpenSetSurvey}, it is currently the state-of-the-art in the open-set recognition problem. In section \ref{sec:method}, we introduce an ID-Conditioned Auto-Encoder (IDCAE), which is the adapted version of C2AE applicable to the unsupervised anomaly detection.

Task 2 in this year IEEE AASP Challenge on Detection and Classification of Acoustic Scenes and Events (DCASE 2020) \cite{dcase2020task2web, dcase2020task2paper} focuses precisely on the unsupervised detection of anomalous sounds for machine condition monitoring. The data for this task is ToyADMOS \cite{ToyASMOS} and MIMII Dataset \cite{MIMII} which consists of sounds of six types of operating machines. In section \ref{sec:model}, we develop the model for this challenge, and we evaluate its performance via an ablation study in section \ref{sec:results}.

\section{Proposed Method}
\label{sec:method}

\begin{figure*}[t]
 \centering
 \centerline{\includegraphics[width=0.7\textwidth]{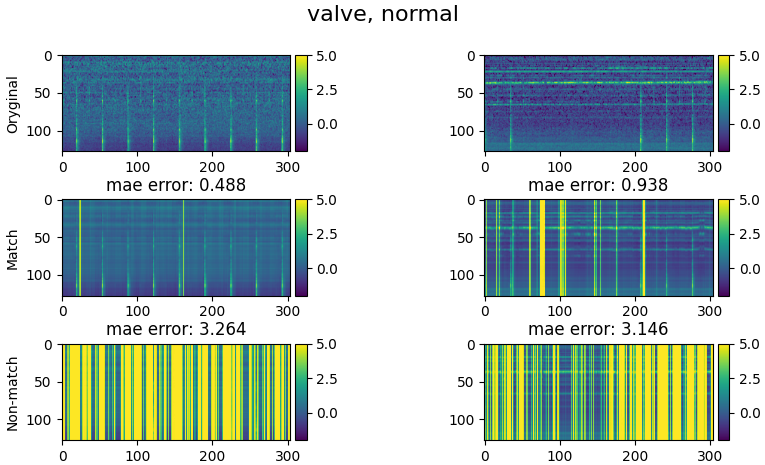}}
 \caption{Two exemplary spectrograms of some valve normal samples with their corresponding reconstructions. In the top row we have original spectrograms of the whole recordings. In the middle row we have reconstructions with matching labels. We see that reconstructions are rather faithful. In the bottom row we have reconstructions with non-matching labels, and as expected the reconstructed spectrograms are quite off.}
 \label{fig:spec_normal}
\end{figure*}

\begin{figure*}[t]
 \centering
 \centerline{\includegraphics[width=0.7\textwidth]{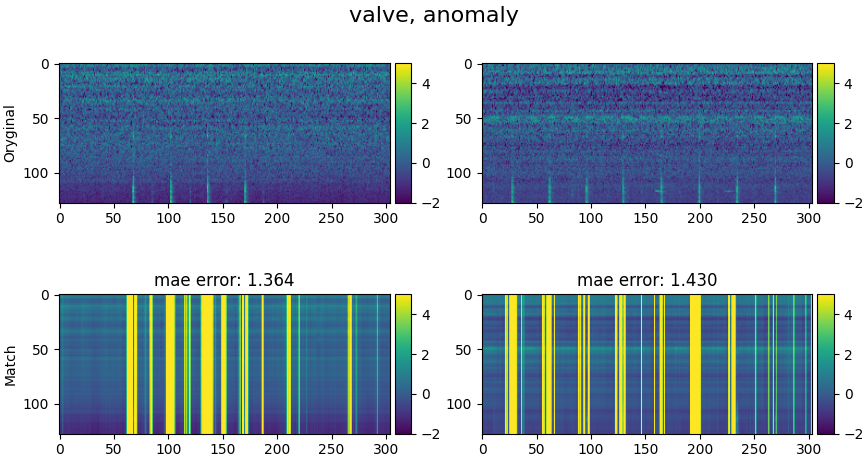}}
 \caption{Two exemplary spectrograms of some valve anomalous samples with their corresponding reconstructions. In the top row, we have original spectrograms of the whole recordings. In the bottom row, we show reconstructed spectrograms with matching labels. The reconstruction error is mostly boosted in the windows containing anomalous valve sounds.}
 \label{fig:spec_anomaly}
\end{figure*}

Let us consider an arbitrary but fixed machine type, henceforth called the machine unless otherwise specified. We assume that we have various IDs of the machine. It is precisely the case in task 2 in the DCASE 2020 Challenge. In the nomenclature from \cite{C2AE} we treat machines with different IDs as distinct classes.

Our system constitutes of three main parts:
\begin{itemize}
  \item encoder $E:\mathcal{X} \to \mathcal{Z}$ which maps \emph{feature vector} $X$ from input space $\mathcal{X}$ to the \emph{code} $E(X)$ in the latent space $\mathcal{Z}$,
  \item decoder $D:\mathcal{Z} \to \mathcal{X}$ which takes the code $Z$ from $\mathcal{Z}$ and outputs the vector $D(Z)$ of the same shape as feature vectors from $\mathcal{X}$,
  \item conditioning made of two functions $H_{\gamma}, H_{\beta}:\mathcal{Y}\to \mathcal{Z}$ which take the one-hot label $l$ from $\mathcal{Y}$ and map it to the vectors $H_{\gamma}(l), H_{\beta}(l)$ of the same size as codes from $\mathcal{Z}$.
\end{itemize}

During feed-forward, the code $Z$ is combined with $H_{\gamma}(l), H_{\beta}(l)$ to form $H(Z,l) = H_{\gamma}(l) \cdot Z + H_{\beta}(l)$, which is an affine transformation of the latent space conditioned by $l$ \cite{Perez2018FiLMVR}. Thus, our whole system takes two inputs $X,l$ from $\mathcal{X}$ and $\mathcal{Y}$ respectively and outputs $D(H(E(X), l))$.

Given an input $X$ with some ID, we call label corresponding to this ID by the \emph{match} and all other labels by \emph{non-matches}. We wish that our system reconstructs $X$ faithfully if and only if it is a normal sample conditioned by the matching label.

Given an input $X$, we set the label $l$ to the match with probability $\alpha$ or to a randomly selected non-match with probability $1-\alpha$, where $\alpha$ is predefined. Thus, for a batch $X_1,X_2, \dots X_n$ approximately $\alpha$ fraction of samples will be conditioned by matches and $1-\alpha$ by non-matches. If $l$ is the match, then the loss equals difference between the system's output and $X$, that is $\norm{D(H(E(X), l)) - X}$. If $l$ is a non-match, then the loss equals difference between the system's output and some predefined constant vector $C$ with the same shape as $X$, that is $\norm{D(H(E(X), l)) - C}$. In our setting $\norm{\cdot}$ is either $L_1$ or the square of $L_2$ norm.

During the inference we always feed the network with matching labels. If a sample is non-anomalous, we expect the reconstruction to be faithful resulting in low reconstruction error (see Figure \ref{fig:spec_normal}). If the sample is anomalous, there may be two cases. If the sample is nothing like any sample during training, then generally auto-encoder wouldn't be able to reconstruct it resulting in high reconstruction error. However, if the sample resemble normal samples with other IDs, then the auto-encoder will try to reconstruct the vector $C$ resulting again in high error (see Figure \ref{fig:spec_anomaly}). 

In general, the point of auto-encoder-based solutions for anomaly detection is to separate distributions of reconstruction errors for normal and anomalous samples. In our case, the distribution of reconstruction errors for anomalies is shifted further to the right due to the samples that resemble samples with other IDs. Thus, it indicates that our method should work at least as good as regular auto-encoder. Overall, distribution of reconstruction errors for anomalies with matching labels should be between distributions of reconstruction errors for normal samples with matching and non-matching labels (see Figure \ref{fig:distribution}).

\begin{figure}[t]
 \centering
 \centerline{\includegraphics[width=0.9\columnwidth]{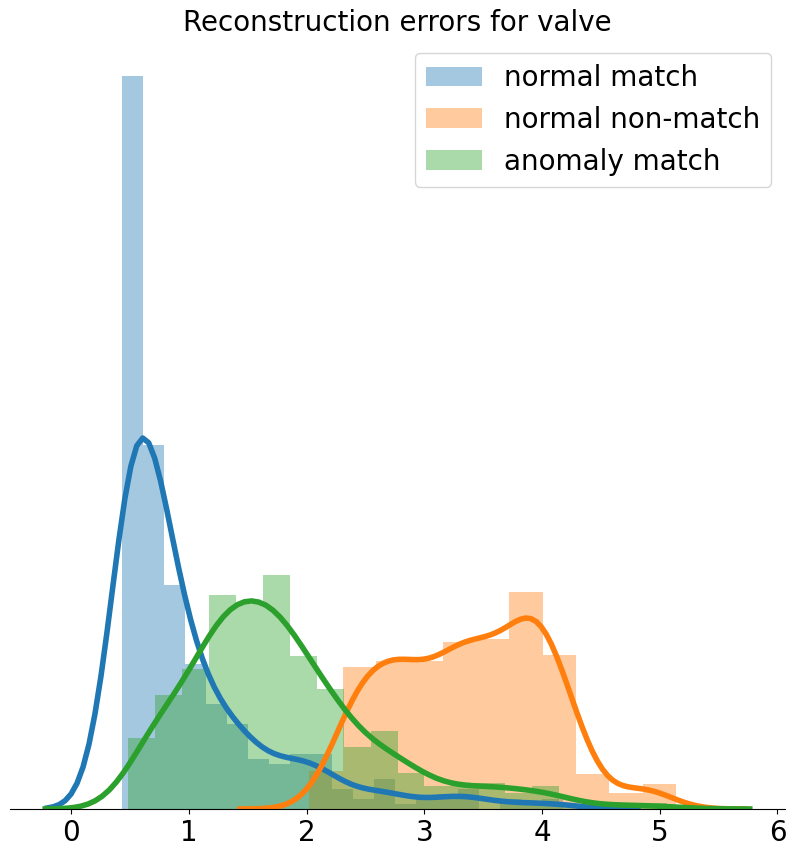}}
 \caption{Distributions of reconstruction errors for all valve samples from the test split from the development dataset.}
 \label{fig:distribution}
\end{figure}

The additional advantage of our approach is that feeding the network with more IDs may result in better performance. In fact, in the section \ref{sec:results} we show that it holds in the case of machine condition monitoring.

\section{Model}
\label{sec:model}

\subsection{Features}
\label{ssec:features}

In our model, we feed the network with fragments of normalised log-mel power spectrograms. Feature vector space $\mathcal{X}$ consists of vectors of the shape $(F, M)$, where $F$ is the frame size and $M$ is the number of mels. Given an audio signal we first compute its Short Time Fourier Transform with 1024 window and 512 hop size, we transform it to the power mel-scaled spectrogram with $M$ mels, and we take its logarithm with base $10$ and multiply it by $10$. Finally, we standardize all spectrograms frequency-wise to zero mean and unit variance, and sample frames of size $F$ as an input to our system.

\subsection{Architecture}
\label{ssec:architecture}

As described in subsection \ref{sec:method}, our model constitutes of the encoder $E$, the decoder $D$ and the conditioning $H_\gamma,H_\beta$. In our case, all these components are fully connected neural networks. Thus, we have to flatten feature vectors and reshape the output to $(F, M)$ for the sake of the dimension compatibility. The dense layers in $E$ and $D$ are followed by batch-norm and relu activation function, while the dense layers in $H_\gamma, H_\beta$ are followed just by sigmoid activation functions. $E$ has three hidden dense layers with $128, 64$ and $32$ units followed by the latent dense layer with $16$ units. $D$ is made of four hidden dense layers each with $128$ units. $H_\gamma$ and $H_\beta$ have both a single hidden dense layer with $16$ units. We summarise the architecture in Table \ref{tab:architecture}.

\begin{table}
  \caption{The architecture of IDCAE}
  \centering
  \begin{tabular}{*3l}
    \toprule
     \textbf{Encoder ($E$)} & \textbf{Decoder ($D$)} & \textbf{Conditioning} ($H_\gamma, H_\beta$) \\
    \midrule
    
    Input $(F,M)$ & Input $16$ & Input \#IDs \\
    Flatten & DenseBlock $128$ & Dense $16$ \\
    DenseBlock $128$ & DenseBlock $128$ & sigmoid \\
    DenseBlock $64$ & DenseBlock $128$ & Dense $16$ \\
    DenseBlock $32$ & DenseBlock $128$ &  \\
    DenseBlock $16$ & Dense $F\cdot M$ &  \\
     & Reshape $(F,M)$ & \\
     \midrule
    \multicolumn{3}{l}{DenseBlock $n$: Dense $n$, Batch-norm, relu} \\
    \bottomrule
  \end{tabular}
  \label{tab:architecture}
\end{table}

\subsection{Parameters}
\label{ssec:parameters}

We used $\alpha=0.75, C=\bm{5}$ (the bold-face denotes that it is a constant vector with value 5 everywhere) with a frame size $F=10$ and number of mels $M=128$. As for complexity for this setup, the encoder, decoder and conditioning has \num[group-separator={,}]{175792}, \num[group-separator={,}]{218880} and \num[group-separator={,}]{800} parameters respectively, making total of \num[group-separator={,}]{395472} parameters.

\subsection{Training}
\label{ssec:training}

We train our models using Adam optimizer with default parameters \cite{Adam} setting mean absolute error as a loss function. For each machine, we train our network for 100 epochs with exponential learning rate decay by multiplying the learning rate by 0.95 every 5 epochs. For every epoch we randomly sample 300 frames from each spectrogram.

\subsection{Ensemble}
\label{ssec:ensemble}

$\alpha, C$ and number of mels $M$ are hyperparamters for our model. Unfortunately, there is no single couple of these parameters that works best for all the machines. We set $F=10$ and did a grid search with $\alpha\in \{0.9, 0.75, 0.5\}, C\in \{\bm{0}, \bm{2.5}, \bm{5}, \bm{10}\}, M\in \{128, 256\}$ trying mean square and mean absolute errors. We selected 3 models for each machine that maximize mAUC (the average of AUC and pAUC with $p=0.1$) on the test split from the development dataset, and for each machine we did an ensemble by selecting 3 weights such that the weighted anomaly score maximize mAUC. The ensemble is illustrated in Figure \ref{fig:ensemble}.

\begin{figure*}[t]
 \centering
 \centerline{\includegraphics[width=0.75\textwidth]{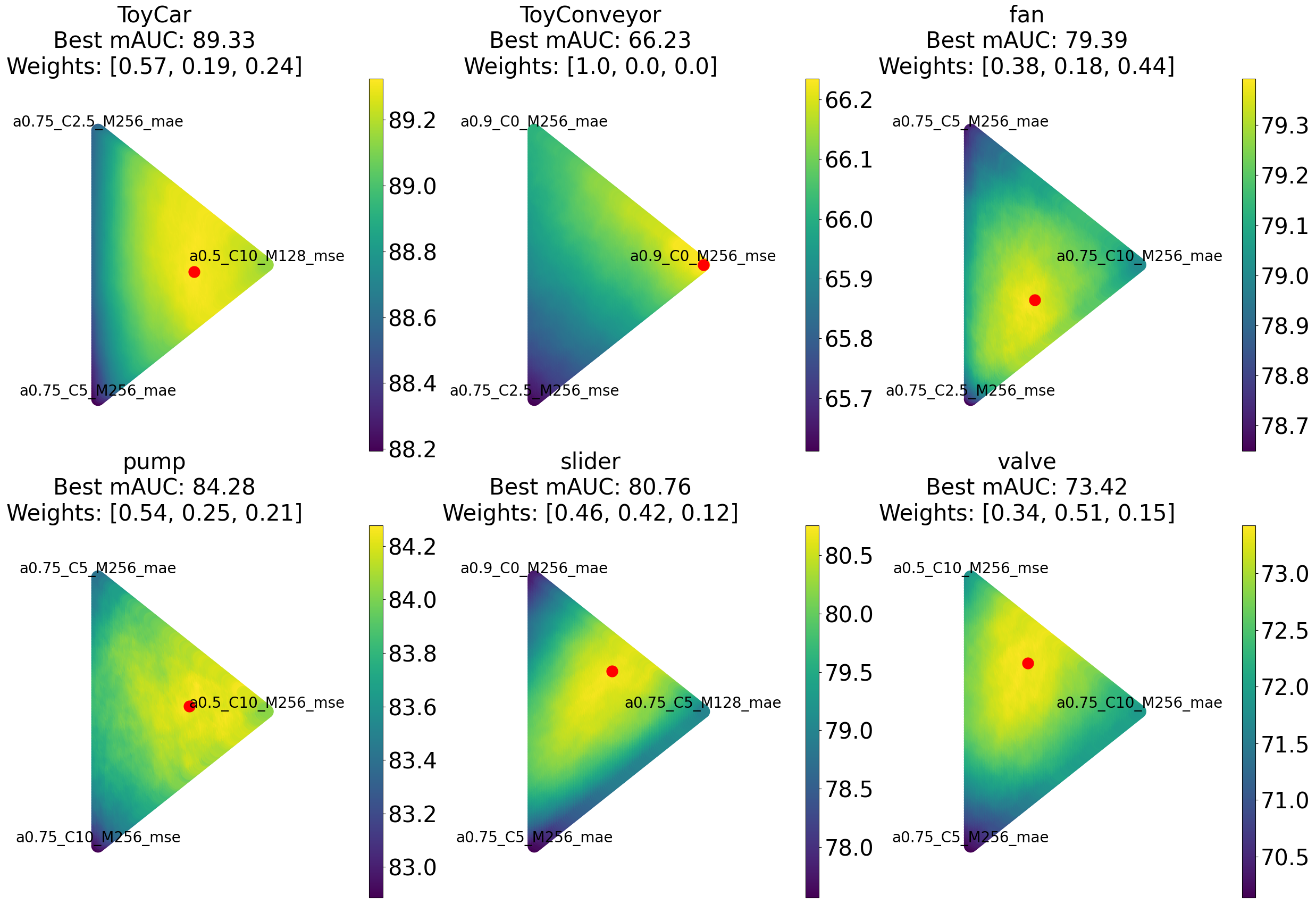}}
 \caption{Visualisation of the ensemble of the best three models for each machine type. The vertices indicate mAUC of individual models and the interior points of triangles indicate mAUC of the convex combination of anomaly scores. That is, for given weights, new anomaly score is obtained as the linear combination of these weights with corresponding anomaly scores, and the mAUC is computed of that scores. The red dots indicate the weights that maximize mAUC.}
 \label{fig:ensemble}
\end{figure*}

\section{Results}
\label{sec:results}

\begin{table*}
\caption{Ablation Study;  the \textbf{boldface} denotes the best scores, and the \underline{underline} denotes the greatest improvements.}
\footnotesize
\centering
\begin{tabular}{l||*2c*2c*2c*2c*2c*2c|*2c}

\toprule
 & \multicolumn{2}{c}{Toy Car} & \multicolumn{2}{c}{Toy Conveyor} & \multicolumn{2}{c}{Fan} & \multicolumn{2}{c}{Pump} & \multicolumn{2}{c}{Slider} & \multicolumn{2}{c|}{Valve} & \multicolumn{2}{c}{Average}\\
 & AUC & pAUC & AUC & pAUC & AUC & pAUC & AUC & pAUC & AUC & pAUC & AUC & pAUC & AUC & pAUC\\
\midrule
\verb|Baseline| & 78.77 & 67.58 & \bf{72.53} & \bf{60.43} & 65.83 & 52.45 & 72.89 & 59.99 & 84.76 & 66.53 & 66.28 & 50.98 & 73.51 & 59.66 \\
\verb|Architect| & 81.18 & 68.78 & 70.37 & 57.21 & 70.63 & 56.44 & 74.15 & \underline{66.89} & 85.68 & 67.92 & 66.47 & 50.79 & 74.75 & 61.34 \\
\verb|Scaler| & 79.09 & 65.26 & 71.23 & 57.38 & 70.51 & 56.32 & 74.23 & 66.08 & 86.28 & 68.33 & 69.46 & 51.65 & 75.13 & 60.84 \\
\verb|Condition| & 78.07 & 74.22 & 70.29 & 59.46 & \underline{77.45} & \underline{70.32} & 77.29 & 70.33 & 80.04 & 68.25 & \underline{78.26} & 55.80 & 76.90 & \underline{66.40} \\
\verb|AddDataset| & \underline{88.66} & \underline{85.54} & 68.26 & 57.37 & 79.56 & 74.21 & \underline{83.28} & 76.75 & 82.50 & 68.76 & 81.97 & 57.05 & \underline{80.71} & 69.95 \\
\verb|Ensemble| & \bf{91.28} & \bf{87.36} & \underline{72.23} & \underline{60.23} & \bf{81.80} & \bf{76.98} & \bf{88.27} & \bf{80.28} & \underline{\bf{86.80}} & \underline{\bf{74.71}} & \bf{84.56} & \underline{\bf{62.28}} & \bf{84.16} & \bf{73.64} \\
\bottomrule

\end{tabular}
\label{tab:ablation}
\end{table*}

We develop and evaluate our system on the dataset from task 2 from DCASE 2020 Challenge \cite{dcase2020task2web}, which consists of recording on 6 different machine types (Toy Car, Toy Conveyor, Fan, Pump, Slider and Valve). The available datasets consists of development and additional datasets. The development dataset consists of 3 or 4 IDs per machine while the additional dataset consist of 3 IDs per machine. All results in this section are evaluated on the test split from the development dataset using the area under the receiver operating characteristic (ROC) curve (AUC) and the partial-AUC (pAUC) with $p=0.1$

We conduct the ablation study starting from the DCASE baseline system and ending on the ensemble. We list the following steps, where each step is built upon the previous one:
\begin{itemize}
    \item \verb|Baseline| - DCASE Baseline system.
    \item \verb|Architect| - Encoder changed to the smaller one with 128, 64 and 32 units, latent layer set to 16 units, frame size set to 10, and training conducted as in subsection \ref{ssec:training}.
    \item \verb|Scaler| - Spectrograms standardised frequency-wise to zero mean and unit variance.
    \item \verb|Condition| - Conditioning added with $\alpha=0.75$ and $C=\bm{5}$.
    \item \verb|AddDataset| - Training on the train splits from both development and additional datasets combining labels from these datasets.
    \item \verb|Ensemble| - Ensemble as described in the subsection \ref{ssec:ensemble}.
\end{itemize}

We summarise the results of the ablation study in Table \ref{tab:ablation}. Even though results vary over machines, we will focus on the average scores which are placed in the last two columns of the Table. Our architecture changes as in \verb|Architect| and normalization \verb|Scaler| have minor influence on AUC and pAUC scores. Adding conditioning layer in \verb|Condition| significantly improves pAUC score, which proves that IDCAE is advantageous over standard AE. Moreover as expected, adding more IDs like in \verb|AddDataset| improves both AUC and pAUC scores significantly. Finally, the ensemble from \verb|Ensemble| due to its nature again significantly boosts both AUC and pAUC.

\section{Conclusion}
\label{sec:conclusion}

In this paper we introduced a method on how to enhance the auto-encoder designed for unsupervised anomaly detection. Assuming distinct IDs, we condition the latent space in order to enforce auto-encoder to reconstruct samples faithfully if and only if they are non-anomalous and conditioned with matching labels. Our method outperforms significantly the DCASE baseline system. In the ablation study, we proved that the conditioning we introduced is crucial for the performance improvement, and that feeding the system with more IDs gets even better results.

\pagebreak

\bibliographystyle{IEEEtran}
\bibliography{refs}
\end{sloppy}
\end{document}